\documentclass[]{spie}  

\usepackage{amsmath,amsfonts,amssymb}
\usepackage{graphicx}
\usepackage[colorlinks=true, allcolors=blue]{hyperref}

\title{Millimeter-wave adaptive optics: Demonstrating closed-loop correction for lowest Zernike modes}

\author[a]{Yoichi Tamura}
\author[b,a]{Akio Taniguchi}
\author[a]{Kotaro Iwakami}
\author[c]{Ichiro Jikuya}
\author[d]{Shion Takeno}
\author[d]{Sachiko K.\ Okumura}
\author[a]{Masaki Sakakibara}
\author[a]{Akinobu Miyake}
\author[a]{Masato Hagimoto}
\author[a]{Kianhong Lee}
\author[a]{Chihiro Imamura}
\author[a]{Sho Fujisawa}
\author[a]{Shutaro Inui}
\author[a]{Masato Kato}
\author[e]{Ryohei Kawabe}
\author[f]{Mikio Kurita}
\author[g]{Nozomi Okada}
\author[a]{Juri Yamanaka}
\affil[a]{Nagoya University, Nagoya, Aichi, 464-8602 Japan}
\affil[b]{Kitami Institute of Technology, Kitami, Hokkaido, 090-8507 Japan}
\affil[c]{Kanazawa University, Kanazawa, Ishikawa, 920-1192 Japan}
\affil[d]{Japan Women's University, Bunkyo, Tokyo, 112-8681 Japan}
\affil[e]{National Astronomical Observatory of Japan, Mitaka, Tokyo, 181-8588 Japan}
\affil[f]{Kyoto University, Kyoto, 606-8501 Japan}
\affil[g]{Osaka Metropolitan University, Osaka, 558-8585 Japan}

\authorinfo{Further author information: (Send correspondence to Y.T.)\\Y.T.: E-mail: ytamura@nagoya-u.jp}

\begin{document} 
\maketitle

\begin{abstract}
We report on a five-element prototype wavefront sensor for millimeter-wave
adaptive optics (MAO), enabling closed-loop correction of tip-tilt and defocus via
secondary mirror (M2) displacement. MAO is essential for large ground-based
millimeter/submillimeter telescopes to maintain surface accuracy under wind and
thermal distortions. Our sensor, based on radio interferometry, measures excess
path lengths from the primary mirror to a focal-plane receiver. A previous 
two-element prototype achieved $< 10~\mu$m accuracy at the Nobeyama 45~m telescope.
The new five-element system, operating at 20~GHz, was installed on the same
telescope. A ``Moon-edge'' experiment confirmed detection of wavefront gradients
through strong correlation with continuum flux. Implementing a PI controller
closed the sensor--M2 loop, stably suppressing the lowest Zernike modes. This
approach establishes a foundation for metrology in future large-aperture
submillimeter facilities such as AtLAST/LST.
\end{abstract}

\keywords{adaptive optics, submillimeter, telescope, metrology, AtLAST, LST}

\section{INTRODUCTION}
\label{sec:intro}  

As ground-based millimeter and submillimeter telescopes
grow in size, maintaining surface accuracy becomes increasingly critical. The mm-wave adaptive optics (MAO) 
addresses this challenge by correcting wind and thermal-induced optical
distortions in real time. Our sensor, based on radio interferometry, measures
variations in excess path lengths from selected points on the primary mirror to a
coherent receiver at the focal plane.\cite{Tamura20} 
A two-element prototype tested at the Nobeyama 45~m telescope achieved less
than 10~$\mu$m accuracy.\cite{Nakano22} The next step is
mapping wavefront errors and correcting them with an active optical component in
closed loop.

To this end, we developed and installed a five-element 20~GHz
prototype on the 45-m telescope, enabling sensing of tip-tilt and defocus Zernike
modes, which can be compensated by lateral and axial sub-reflector (M2) displacements. We
first conducted a ``Moon-edge'' experiment, monitoring wavefront gradients
(pointing errors) in the lunar radial direction. Simultaneous 20~GHz continuum flux
measurements reflected beam filling factors and showed strong correlation with
sensor outputs, indicating detection of the wavefront gradient. Finally, we
implemented a PI controller to close the sensor--M2 loop\cite{Jikuya26} (Jikuya et al.\ 2026, Proc.
SPIE, this volume), demonstrating stable suppression of the lowest Zernike
modes. This capability provides a foundation for metrology in future large-aperture
submillimeter telescopes such as AtLAST/LST.\cite{Mroczkowski25, Kawabe16}

\begin{figure} [ht]
\begin{center}
   \includegraphics[width=0.9 \textwidth]{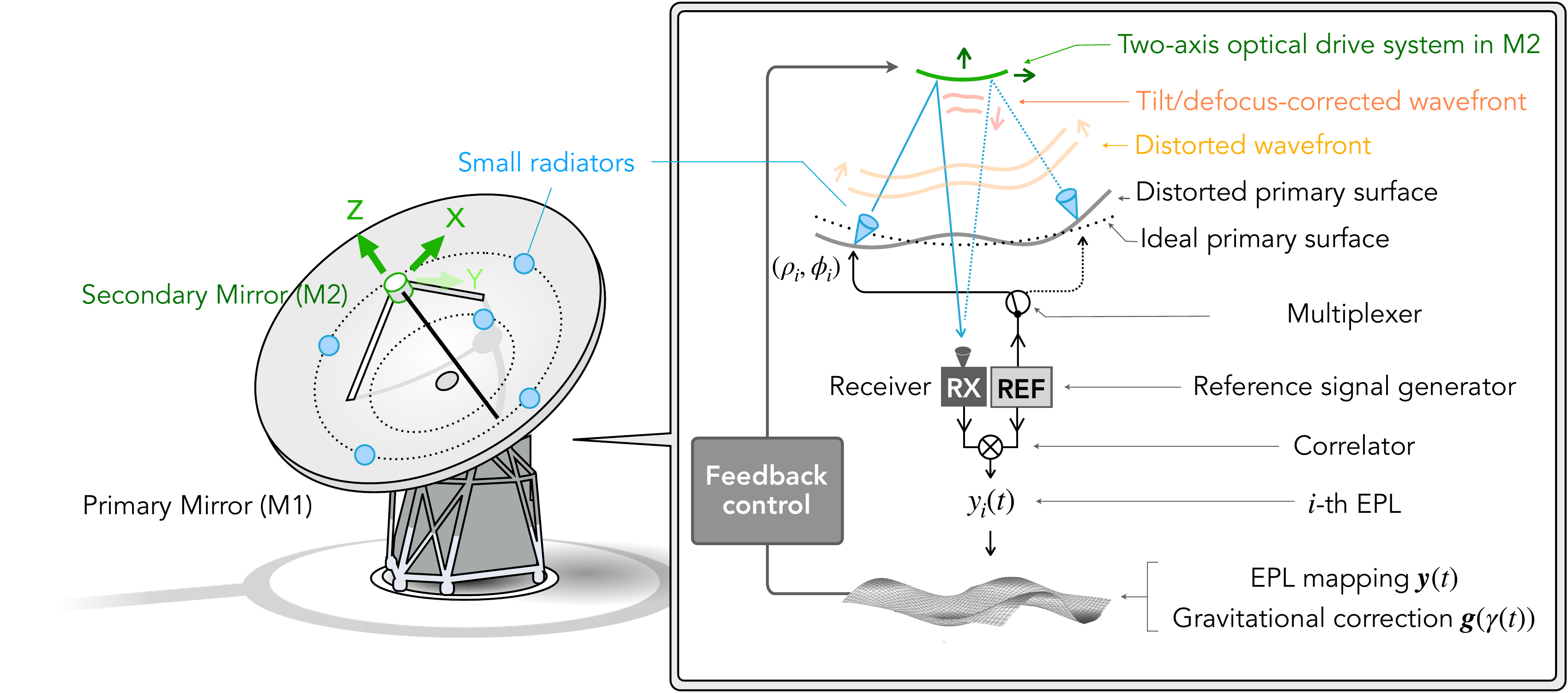}
\end{center}
\caption[example] 
{ \label{fig:concept} 
Concept of a five-element wavefront sensor and closed-loop correction for lowest Zernike modes tested at the Nobeyama 45~m telescope. 
A reference signal is radiated from small transmitters mounted on the primary mirror (M1) and received at the focal plane through the telescope optics. 
The correlator measures the excess path length (EPL) at each sampled position, producing an EPL map across the aperture. 
After correction for the EPL component of elevation-dependent gravitational deformation, the measured low-order wavefront error is fed back to the two-axis secondary mirror (M2), enabling real-time compensation of tilt- and defocus-like modes.}
\end{figure}

\section{FIVE-ELEMENT WAVEFRONT SENSOR}
To extend our previous two-element demonstration\cite{Nakano22} to low-order wavefront mapping,
we developed a five-element prototype sensor operating at 20~GHz and installed it
on the Nobeyama 45~m telescope. The system is based on aperture-plane
interferometry:\cite{Tamura20} a broadband reference signal generated by a noise source is transferred through radio-over-fiber and an optical switch to the primary mirror (M1), radiated from small feed horns
mounted at five locations on M1, and received by the focal-plane
H22 receiver through the normal telescope optics. By correlating the received signal
with the reference signal, we measure the excess path length (EPL) at each feed
position. The five feeds were arranged in a cross-like pattern (center, top, right,
bottom, and left), providing sensitivity to the lowest spatial modes of the
wavefront, in particular tip-tilt and defocus. 
This configuration was chosen as the minimum system
capable of detecting and separating the lowest Zernike-like modes that can be
compensated by axial and lateral displacement of M2.
In the present implementation, stability is on timescales of $\sim 30$~s, which is only sufficient for probing the low-frequency deformations expected from wind and thermal effects on that timescale. 


\begin{figure} [ht]
\begin{center}
   \includegraphics[width=\textwidth]{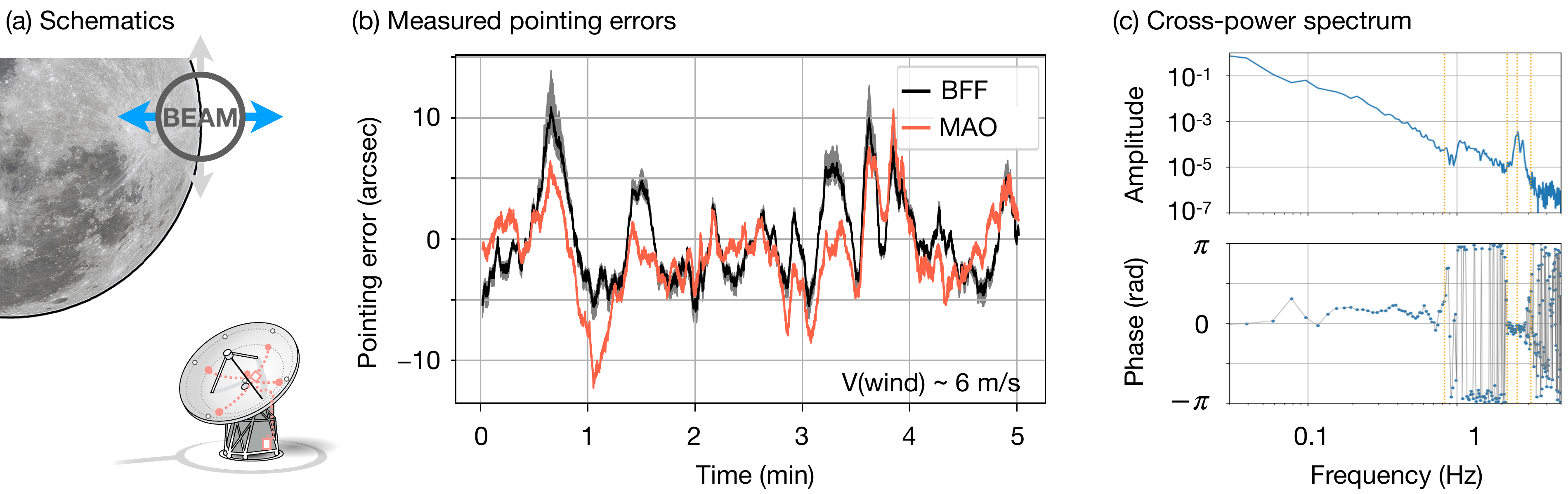}
\end{center}
\caption[example] 
{ \label{fig:moon} 
Moon-edge experiment demonstrating sensitivity of the five-element MAO sensor to the lowest-order wavefront gradient. 
(a) Schematic of the observation at the lunar limb, where radial pointing fluctuations change the beam filling factor (BFF) measured by the receiver total power. 
(b) Time series of pointing variations derived from the BFF and from the MAO wavefront sensor, showing a clear correlation under a wind speed of $\sim 6$~m~s$^{-1}$. 
(c) Cross-power spectrum between the two signals, indicating that the dominant correlated component lies at low frequencies $\lesssim 1$~Hz.}
\end{figure} 

\section{MOON-EDGE EXPERIMENT}
As a first on-sky validation of the five-element sensor, we carried out a
``Moon-edge'' experiment at 20~GHz (Fig.~\ref{fig:moon}a). In this test, the telescope beam was placed on
the lunar limb so that a small radial pointing fluctuation changes the beam filling
factor (BFF) and therefore the continuum total power measured by the receiver. At the same
time, the MAO sensor measured differential EPLs across the aperture. Because the
dominant first-order wavefront deformation appears as a wavefront gradient, the
sensor output should correlate with the pointing error inferred from the BFF 
if the sensor is responding to the expected low-order mode.

We found a clear positive correlation between the wavefront gradient estimated by
the five-element MAO sensor and the pointing variation traced by the continuum
signal (Fig.~\ref{fig:moon}b). The variations were dominated by low temporal frequencies, typically $\lesssim 1$~Hz (Fig.~\ref{fig:moon}c), 
consistent with the slowly varying nature of large-scale structural
deformation in the telescope. Although some residual difference remained between the
two observables, plausibly due to higher-order modes such as defocus or astigmatism
that are not captured by a simple one-dimensional beam-filling estimate or anomalous refraction caused by the atmospheric water vapor layer that the wavefront sensor does not probe, the
experiment verified that the sensor successfully detects the lowest-order wavefront
gradient in real observing conditions.

\begin{figure} [ht]
\begin{center}
   \includegraphics[width=\textwidth]{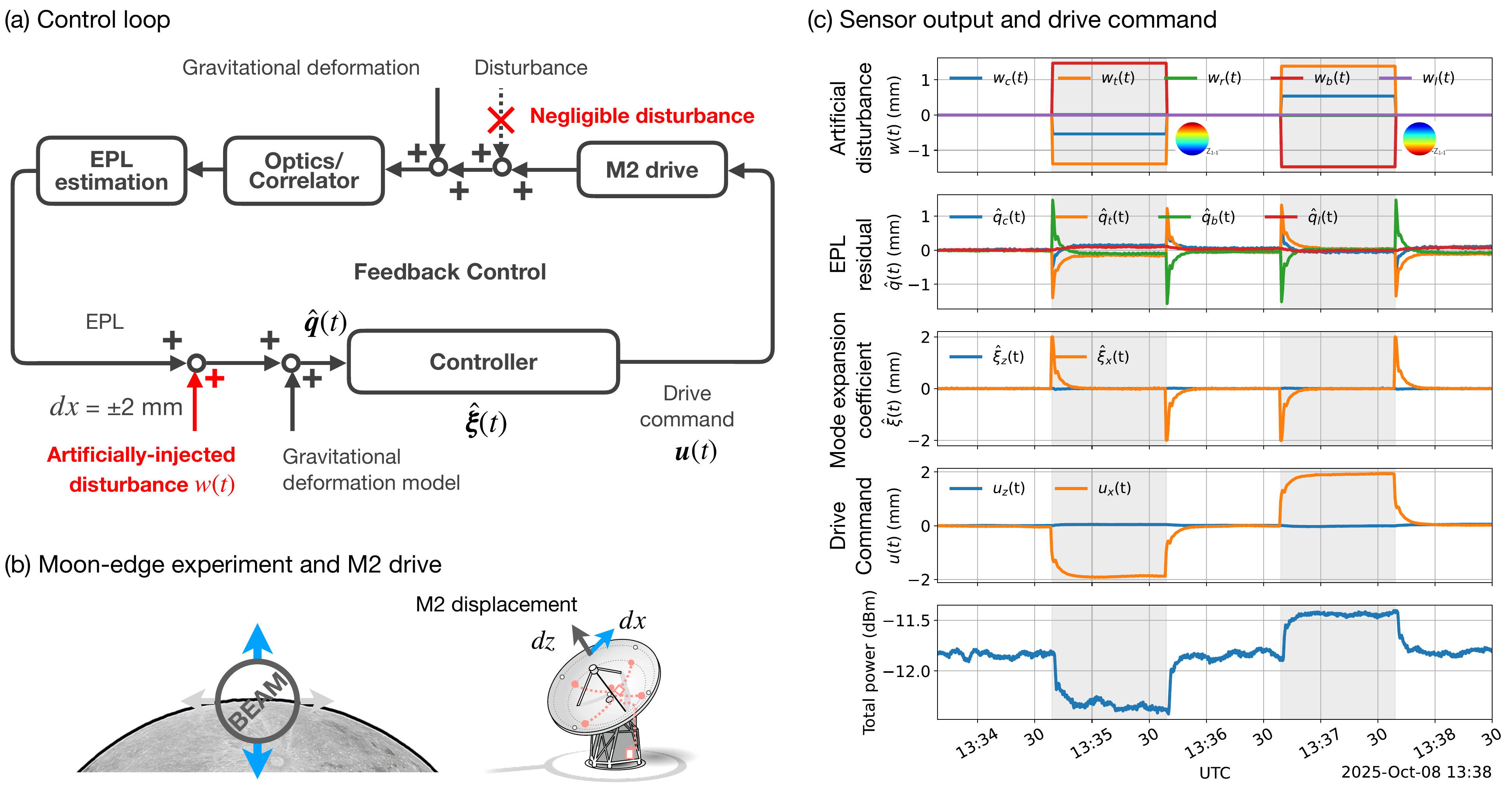}
\end{center}
\caption[example] 
{ \label{fig:loop} 
Closed-loop correction of a low-order wavefront mode using the MAO sensor and M2 drive. 
(a) Block diagram of the feedback system. The measured excess path length (EPL) is corrected for the elevation-dependent gravitational contribution ($\hat{\boldsymbol{q}}(t)$), projected onto the controllable modal subspace $\hat{\boldsymbol{\xi}}(t)$, and converted into M2 drive commands $\boldsymbol u(t)$ by the control law. 
(b) The tilt-like beam response to M2 displacement in the $dx$ direction. 
(c) Time series for a disturbance corresponding to the $Z_{1}^{-1}$ tilt-like mode. From top to bottom, the panels show the artificially-injected disturbance vector $\boldsymbol w(t)$, the residual EPL vector $\hat{\boldsymbol q}(t)$, the modal expansion coefficient vector $\hat{\boldsymbol\xi}(t)$, the drive command ${\boldsymbol u}(t)$, and the receiver power during the Moon-edge observation. The shaded intervals indicate injected disturbances. The controller produces the expected M2 response and suppresses the controllable low-order mode in a stable manner.
}
\end{figure} 

\section{CLOSED-LOOP CORRECTION}
The next step was to close the loop between the wavefront sensor and the
M2 drive. Since the five-element configuration mainly probes the lowest
wavefront modes, we focused on the components that can be generated and corrected by
M2 motion. In the present telescope system, axial motion of M2 primarily produces a
defocus-like mode, while lateral motion along one available axis produces a
tilt-like mode. The relation between M2 displacement and the five measured EPLs was
characterized by a measurement matrix obtained from dedicated calibration
experiments. After subtracting the elevation-dependent contribution of gravitational
deformation, the residual EPL vector was projected onto the controllable low-order
subspace and fed to a proportional-integral (PI) controller\cite{Jikuya26} (Fig.~\ref{fig:loop}a, see Jikuya et al.\ 2026, Proc.\ SPIE, this volume).

Closed-loop experiments demonstrated stable feedback operation and suppression of
artificially injected disturbances in the controllable modes. When a disturbance
equivalent to a correctable low-order wavefront error was applied (Fig.~\ref{fig:loop}b, $dx = \pm2$~mm), the estimated
mode amplitudes $\hat{\boldsymbol\xi}(t)$ and residual EPLs $\hat{\boldsymbol q}(t)$ converged toward zero, and the receiver response
changed consistently with cancellation by the M2 drive command $\boldsymbol u(t)$ (Fig.~\ref{fig:loop}c). In contrast, when
a disturbance dominated by an uncontrollable mode was applied, the controller did
not significantly react, as expected from the limited actuator space. These results
show that the sensor--M2 loop can stably suppress the lowest Zernike modes accessible
to the current hardware, establishing the first demonstration of closed-loop
wavefront compensation at the Nobeyama 45~m telescope.


\acknowledgments 
 
We thank A.\ Kiselev, S.\ Thoms, T.\ Mroczkowski, and K.\ Kimura for fruitful discussions. We also thank K.\ Handa, C.\ Miyazawa, T.\ Kurakami, and A.\ Nishimura at Nobeyama Radio Observatory for their support. This study is financially supported by JSPS KAKENHI (No.\ 17H06206, 22H04939, 23K20035, 24H01808) and NAOJ Research Coordination Committee, NINS.

\bibliography{report} 
\bibliographystyle{spiebib} 

\end{document}